\documentstyle [12pt] {article}
\title{QUANTUM, PHOTO-ELECTRIC SINGLE CAPACITOR PARADOX}

\author{Darko V. Kapor, Vladan Pankovi\'c\\
Department of Physics, Faculty of Sciences, 21000 Novi Sad,\\ Trg
Dositeja Obradovi\'ca 4., Serbia, vpankovic@if.ns.ac.yu}

\date {}
\begin{document}
\maketitle \vspace {0.5cm}
 \vspace {0.1cm}

\begin {abstract}
In this work single capacitor paradox (a variation of the
remarkable two capacitor paradox) is considered in a new, quantum
discrete form. Simply speaking we consider well-known usual,
photoelectric effect experimental device, i.e. photo electric
cell, where cathode and anode are equivalently charged but
non-connected. It, obviously, represents a capacitor that
initially, i.e. before action of the photons with individual
energy equivalent to work function, holds corresponding energy of
the electrical fields between cathode and anode. Further, we
direct quantum discretely photons, one by one, toward cathode
where according to photo-electrical effect electrons discretely,
one by one, will be emitted and directed toward anode. It causes
discrete discharge of the cell, i.e. capacitor and discrete
decrease of the electrical field. Finally, total discharge of the
cell, i.e. capacitor, and total disappearance of the electrical
field and its energy will occur. Given, seemingly paradoxical,
capacitor total energy loss can be simply explained without any
dissipative effects (Joule heating or electromagnetic waves
emission can be neglected as high order small corrections) by work
done by the electrical field by movement of the electrons from
cathode to anode. (Remarkable two capacitors paradox can be,
obviously, formulated and explained in the completely analogous
way.)
\end {abstract}

\vspace {1cm}

As it is well-known remarkable two-capacitors paradox, formulated
and considered in many textbooks and articles on the basic
principles and applications of the electronic and electrodynamics
[1]-[7], states the following. Consider an ideal (without any
electrical resistance and inductivity) electrical circuit with
first, initially charged, and second, initially non-charged, of
two identical capacitors. In given circuit, by transition from
initial, open state (switch OFF state) in the closed state (switch
ON state), an unexpected, mysterious loss of the half of initial
energy of electrical fields within capacitors occurs. Different
authors [4]-[7] suggest that given energy loss is realized by
different dissipative processes (Joule heating or/and
electromagnetic waves emissions) realized by non-neglectable
residual electric resistances and inductivities in realistic
circuits.

In this work single capacitor paradox (a variation of the
remarkable two capacitor paradox) will be considered in a new,
discrete form. Simply speaking we shall consider well-known usual,
photoelectric effect experimental device, i.e. photo electric
cell, where cathode and anode are equivalently charged but
non-connected. It, obviously, represents a capacitor that
initially, i.e. before action of the photons with individual
energy equivalent to work function, holds corresponding energy of
the electrical fields between cathode and anode. Further, we shall
direct quantum discretely photons, one by one, toward cathode
where according to photo-electrical effect electrons discretely,
one by one, will be emitted and directed toward anode. It causes
discrete discharge of the cell, i.e. capacitor and discrete
decrease of the electrical field. Finally, total discharge of the
cell, i.e. capacitor, and total disappearance of the electrical
field and its energy will occur. Given, seemingly paradoxical,
total energy loss can be simply explained without any dissipative
effects (Joule heating or electromagnetic waves emission can be
neglected as high order small corrections) by work done by the
electrical field by movement of the electrons from cathode to
anode. (Remarkable two capacitors paradox can be, obviously,
formulated and explained in the completely analogous way.)

Consider well-known usual, photoelectric effect experimental
device, i.e. photo-electric cell, where cathode and anode are
non-connected and equivalently charged by  $Q=Ne$ charge (which,
of course, means that cathode holds N electrons, while at the
anode N electrons are missing.) Obviously, given cell can be
considered as a charged capacitor with some constant capacity C
and voltage $V=\frac {Q}{C}=\frac {Ne}{C}=Nv$ for elementary
voltage $v=\frac {e}{C}$.

Electric field of the capacitor holds initially, before any action
of the light at the cathode, the following energy
\begin {equation}
 E_{in}= \frac { CV^{2}}{2} = \frac {Q^{2}}{2C} =  \frac {N^{2}e^{2}}{2C}= \frac {QV}{2}
\end {equation}

Further, we shall direct quantum discretely photons, one by one,
from an external source toward cathode where according to
photo-electrical effect electrons discretely, one by one, will be
emitted and directed toward anode. It causes discrete discharge of
the cell, i.e. capacitor and discrete decrease of the electrical
field. Finally, total discharge of the cell, i.e. capacitor, and
total disappearance of the electrical field and its energy will
occur. Then, energy of the electric field within cell, i.e.
capacitor becomes zero, i.e.
\begin {equation}
 E_{fin}= 0         .
\end {equation}

In this way there is the following energy difference, i.e.
diminishing between the initial and final state of  given cell,
i.e. capacitor
\begin {equation}
 \Delta E = E_{fin} - E_{in}= - E_{in} = - \frac {QV}{2} .
\end {equation}
It seems as a paradoxical total energy loss.

Consider now energy decrease of the electrical field of photo
cell, i.e. capacitor more accurately.

By photo emission of the first electron, total energy of the
photon is compensated by work function. It implies that first
photo electron obtains finally kinetic energy by electrical field
only and this energy equals eV. We shall consider only such
"idealized" cases when given kinetic energy is much smaller than
working function. In given cases really existing dissipative
processes can be neglected as high order small corrections.

When given electron interact with anode it yields its kinetic
energy to anode and becomes captured. It causes diminishing of the
charge of initial capacitor for e so that final capacitor holds
charge $(N-1)e$, voltage $(N-1)e/C$ and electrical field energy
\begin {equation}
  E_{1}= \frac {(N-1)^{2}}{2C}\simeq \frac {(Ne)^{2}}{2C} - \frac {Ne^{2}}{C}= E - eV  \hspace{1cm} {\rm for} \hspace{0.5 cm} N\gg1.
\end {equation}
It implies that diminishing of the energy of the electric field
equals
\begin {equation}
   \Delta E_{1}= E_{1}- E_{in}\simeq  - eV = -eNv    \hspace{1cm} {\rm for} \hspace{0.5 cm}  N \gg 1
\end {equation}
corresponds to work eV done by  electrical field by movement of
the one electron from one in the other capacitor plate, i.e. from
cathode to anode.

Simple induction implies that total work, done by discretely
diminishing of the electric field energy by moving of all N
electrons from cathode to anode, equals
\begin {equation}
   W =  -ev (N + (N-1) + … + 1) = ev\frac {N(N-1)}{2} \simeq N^{2}\frac{ev}{2}= \frac {(Ne)( Nv)}{2} = \frac {QV}{2} .
\end {equation}

In this way we obtain very simple and reasonable solution of given
quantum discrete, single capacitor paradox in the completely ideal
(without any electrical resistance or inductivity) electrical
circuit. (More precisely we shall consider that really existing
resistance and inductivities yield only high order corrections
which here can be neglected.)

In conclusion, the following can be shortly repeated and pointed
out. Simply speaking we consider well-known usual, photo-electric
effect experimental device, i.e. photo-electric cell, where
cathode and anode are equivalently charged but non-connected. It,
obviously, represents a capacitor that initially, i.e. before
action of the photons with individual energy equivalent to work
function, holds corresponding energy of the electrical fields
between cathode and anode. Further, we direct quantum discretely
photons, one by one, toward cathode where according to
photo-electric effect electrons discretely, one by one, become
emitted and directed toward anode. It causes discrete discharge of
the cell, i.e. capacitor and discrete decrease of the electrical
field. Finally, total discharge of the cell, i.e. capacitor, and
total disappearance of the electrical field and its energy will
occur. Given, seemingly paradoxical, capacitor total energy loss
can be simply explained without any dissipative effects (Joule
heating or electromagnetic waves emission can be neglected as high
order small corrections) by work done by the electrical field by
movement of the electrons from cathode to anode. (Remarkable two
capacitors paradox can be, obviously, formulated and explained in
the completely analogous way.)

\vspace{0.5cm}

     Authors are deeply grateful to Prof. Dr. Tristan H$\ddot {\rm u}$bsch for illuminating discussions.

\vspace{0.5cm}

 {\large \bf References}

\begin {itemize}

\item [[1]] D. Halliday, R. Resnick, {\it Physics, Vol. II} (J. Willey, New York, 1978)
\item [[2]] F. W. Sears, M.W. Zemansky, {\it University Physics} (Addison-Wesley, Reading, MA, 1964)
\item [[3]] M. A. Plonus, {\it Applied Electromagnetics}, (McGraw-Hill, New York, 1978)
\item [[4]] E. M. Purcell, {\it Electricity and Magnetism, Berkeley Physics Course Vol. II} (McGraw-Hill, New York, 1965)
\item [[5]] R. A. Powel, {\it Two-capacitor problem: A more realistic view}, Am. J. Phys. {\bf 47} (1979) 460
\item [[6]] T. B. Boykin, D. Hite, N. Singh, Am. J. Phys. {\bf 70} (2002) 460
\item [[7]] K. T. McDonald, {\it A Capacitor Paradox}, class-ph/0312031

\end {itemize}

\end {document}